\documentclass[english,manuscript,A4,Engish,superscriptaddress]{revtex4-1}
\usepackage[T1]{fontenc}
\usepackage[latin9]{inputenc}
\usepackage{color}
\usepackage{babel}
\usepackage{amsbsy}
\usepackage{amstext}
\usepackage{graphicx}
\usepackage{esint}
\usepackage[unicode=true,pdfusetitle,
 bookmarks=true,bookmarksnumbered=false,bookmarksopen=false,
 breaklinks=false,pdfborder={0 0 1},backref=false,colorlinks=true]
 {hyperref}
\hypersetup{
 pdfborderstyle=,linkcolor=blue,citecolor=cyan}

\makeatletter

\providecommand{\tabularnewline}{\\}

\usepackage{babel}

\usepackage{xcolor}

\makeatother

\begin{document}
\title{Photoproduction of $e^{+}e^{-}$ in peripheral isobar collisions }
\author{Shuo Lin}
\email{linshuo@mail.ustc.edu.cn}

\affiliation{Department of Modern Physics, University of Science and Technology
of China, Anhui 230026, China}
\author{Ren-Jie Wang}
\email{wrjn@mail.ustc.edu.cn}

\affiliation{Department of Modern Physics, University of Science and Technology
of China, Anhui 230026, China}
\author{Jian-Fei Wang}
\email{wjf1996@mail.ustc.edu.cn}

\affiliation{Department of Modern Physics, University of Science and Technology
of China, Anhui 230026, China}
\author{Hao-Jie Xu}
\email{haojiexu@zjhu.edu.cn}

\affiliation{School of Science, Huzhou University, Huzhou, Zhejiang, 313000, China}
\author{Shi Pu}
\email{shipu@ustc.edu.cn}

\affiliation{Department of Modern Physics, University of Science and Technology
of China, Anhui 230026, China}
\author{Qun Wang}
\email{qunwang@ustc.edu.cn}

\affiliation{Department of Modern Physics, University of Science and Technology
of China, Anhui 230026, China}
\begin{abstract}
We investigate the photoproduction of di-electrons in peripheral collisions
of $\ensuremath{_{44}^{96}}\textrm{Ru}+\ensuremath{_{44}^{96}}\textrm{Ru}$
and $\ensuremath{_{40}^{96}}\textrm{Zr}+\ensuremath{_{40}^{96}}\textrm{Zr}$
at 200 GeV. With the charge and mass density distributions given by
the calculation of the density functional theory, we calculate the
spectra of transverse momentum, invariant mass and azimuthal angle
for di-electrons at 40-80\% centrality. The ratios of these spectra
in Ru+Ru collisions over to Zr+Zr collisions are shown to be smaller
than $(44/40)^{4}$ (the ratio of $Z^{4}$ for Ru and Zr) at low transverse
momentum. The deviation arises from the different mass and charge
density distributions in Ru and Zr. So the photoproduction of di-leptons
in isobar collisions may provide a new way to probe the nuclear structure.
\end{abstract}
\maketitle

\section{Introduction}

In ultra-relativistic heavy ion collisions, extremely strong electromagnetic
fields (of order $10^{15}$ Tesla) are generated when two colliding
nuclei pass through each other~\citep{Skokov:2009qp,Bzdak:2011yy,Deng:2012pc,Roy:2015coa,Pu:2016ayh}.
Such strong electromagnetic fields provide an experimental platform
for the study of novel quantum transport phenomena under extreme conditions,
such as the chiral magnetic and separation effects~\citep{Kharzeev:2007jp,Fukushima:2008xe},
the chiral electric separation effect~\citep{Huang:2013iia,Pu:2014cwa},
and other nonlinear effects ~\citep{Pu:2014fva,Chen:2016xtg,Hidaka:2017auj}.
These chiral transport phenomena can be described by microscopic quantum
kinetic theories~\citep{Gao:2019znl,Weickgenannt:2019dks,Weickgenannt:2020aaf,Hattori:2019ahi,Wang:2019moi,Yang:2020hri,Weickgenannt:2020sit,Li:2019qkf,Liu:2020flb,Weickgenannt:2021cuo,Wang:2021qnt,Sheng:2021kfc,Huang:2020wrr,Stephanov:2012ki,Son:2012zy,Gao:2012ix,Chen:2012ca,Manuel:2013zaa,Manuel:2014dza,Chen:2014cla,Chen:2015gta,Chen:2013dca,Hidaka:2016yjf,Hidaka:2017auj,Mueller:2017lzw,Hidaka:2018ekt,Hidaka:2018mel,Gao:2018wmr,Huang:2018wdl,Liu:2018xip,Lin:2019ytz,Lin:2019fqo,Yamamoto:2020zrs,Hidaka:2022dmn,Fang:2022ttm,Gao:2019zhk,Gao:2017gfq,Luo:2021uog}
and macroscopic magnetohydrodynamics~\citep{Roy:2015kma,Pu:2016ayh,Siddique:2019gqh,Shi:2017cpu,Pu:2016bxy,Inghirami:2016iru},
see, e.g., Refs. \citep{Kharzeev:2015znc,Liao:2014ava,Miransky:2015ava,Huang:2015oca,Fukushima:2018grm,Bzdak:2019pkr,Zhao:2019hta,Gao:2020vbh,Hidaka:2022dmn}
for recent reviews. On the other hand, it is also possible to study
nonlinear effects of quantum electrodynamics (QED) in ultra-relativistic
heavy ion collisions, such as light-by-light scatterings~\citep{ATLAS:2017fur},
matter generation directly from photons~\citep{STAR:2019wlg,Zha:2018tlq},
vacuum birefringence~\citep{Hattori:2012je,Hattori:2012ny,Hattori:2020htm,STAR:2019wlg,Hattori:2022uzp,Adler:1971wn}
and Schwinger mechanism~\citep{Schwinger:1951nm,Copinger:2018ftr,Copinger:2020nyx,Copinger:2022jgg}.

In recent years, the lepton pair photoproduction in peripheral and
ultra-peripheral collisions has been extensively studied in both experiments
and theories. To give a better understanding of experimental data~\citep{ATLAS:2018pfw,STAR:2018ldd,STAR:2019wlg,ALICE:2022hvk},
besides the equivalent photon approximation (EPA) by STARlight~\citep{Klein:2016yzr},
several theoretical methods have been developed, such as QED models
with generalized EPA in the background field approach~\citep{Vidovic:1992ik,Hencken:1994my,Hencken:2004td,Zha:2018tlq,Zha:2018ywo,Brandenburg:2020ozx,Brandenburg:2021lnj,Li:2019sin,Wang:2022ihj},
the method based on the factorization theorem~\citep{Klein:2018fmp,Klein:2020jom,Li:2019yzy,Xiao:2020ddm}
and the QED model with the wave-packet description of nuclei~\citep{Wang:2021kxm,Wang:2022gkd}.
Furthermore, it has been shown in Ref.~\citep{Li:2019yzy,Li:2019sin}
that linearly polarized photons, similar to linearly polarized gluons~\citep{Akcakaya:2012si,Schafer:2012yx,Boer:2017xpy,Dominguez:2011br,Metz:2011wb},
can generate the azimuthal angle modulation measured by the STAR collaboration~\citep{STAR:2019wlg}.
A similar azimuthal angle asymmetry in diffractive production of pions
related to elliptic gluon Wigner distribution in ultra-peripheral
collisions is proposed in Ref.~\citep{Hagiwara:2021qev}. So the
photonuclear reaction can be used to probe the properties of initial
gluons~\citep{Xing:2020hwh,Hagiwara:2021qev,STAR:2022wfe,STAR:2021wwq,Brandenburg:2022jgr}.

The isobar collisions of $\ensuremath{_{44}^{96}}\textrm{Ru}+\ensuremath{_{44}^{96}}\textrm{Ru}$
and $\ensuremath{_{40}^{96}}\textrm{Zr}+\ensuremath{_{40}^{96}}\textrm{Zr}$
at the top collision energy of RHIC were originally proposed to search
for the chiral magnetic effect (CME) \citep{Voloshin:2010ut}. Since
Ru and Zr are isobars (with the same nucleon number but different
proton numbers), the electromagnetic fields and thus the chiral magnetic
effect should be different in isobar collisions at the same centrality,
while the backgrounds related to collision geometry, such as the elliptic
flow $v_{2}$ and charged hadron multiplicity $N_{{\rm ch}}$, are
expected to be the same. However, according to the calculation based
on the energy density functional theory (DFT), there are sizable differences
in nuclear density distributions for CME backgrounds which ruin the
initial premise of isobar collisions for the CME search~\citep{Xu:2017zcn,Li:2018oec}.
This has recently been confirmed by the isobar data of STAR collaboration~\citep{STAR:2021mii},
indicating that the structure of isobar nuclei is crucial to the baseline
for the CME signal. The lepton pair photoproduction depends on the
charge distributions of colliding nuclei, which may provide a further
constraint on the nuclear structure parameters in isobar collisions.


In this paper, we employ the theoretical method developed in previous
studies~\citep{Wang:2021kxm,Wang:2022gkd} by some of us to investigate
the lepton pair photoproduction in isobar collisions. The method is
based on QED in a classical field approximation with the wave-packet
description of colliding nuclei encoding the information of the polarization
and transverse momentum (or impact parameter) dependence of photons
in the differential cross section. It can describe the photoproduction
data of lepton pairs in peripheral and ultra-peripheral collisions
\citep{STAR:2018ldd,STAR:2019wlg,Zhou:2022gbh}.

We will calculate the transverse momentum, invariant mass and azimuthal
angle distributions for $e^{+}e^{-}$ pairs at $\sqrt{s_{_{{\rm NN}}}}=200$
GeV in Ru+Ru and Zr+Zr collisions. The Woods-Saxon parameters for
isobar nuclei are obtained by the state-of-art DFT calculation through
nuclear charge and mass density distributions \citep{Xu:2021qjw}.
Since the nuclear mass and charge density distributions give sizable
difference in multiplicity distributions in isobar collisions, the
lepton pair photoproduction is calculated with the charge density
distribution, while the centrality is defined from the Glauber model
with the nuclear mass density. The centrality determined from the
charge density distribution will also be computed as a control. This
study provides a new way of probing the nuclear structure through
photoproduction of lepton pairs in isobar collisions.


The paper is organized as follows. In Sec.~\ref{sec:Theoretical},
we briefly review the theoretical method \citep{Wang:2021kxm} and
introduce the parameters in the numerical calculation. In Sec.~\ref{sec:distriubtion},
we present the transverse momentum, invariant mass and azimuthal angle
distributions for $e^{+}e^{-}$ at $\sqrt{s_{_{{\rm NN}}}}=$200 GeV
in Ru+Ru and Zr+Zr collisions. We study the charge and centrality
dependence of the cross section in Sec.~\ref{sec:Charge-and-centrality}.
We make a summary of the main result in Sec.~\ref{sec:Conclusion}.

\textit{Notational convention}. We use $\mathbf{P}_{{\rm T}}^{\mathrm{ee}}=\mathbf{k}_{{\rm 1T}}+\mathbf{k}_{{\rm 2T}}$
for the transverse momentum of the electron pair and $\mathbf{K}_{{\rm T}}^{\mathrm{ee}}=\frac{1}{2}(\mathbf{k}_{{\rm 2T}}-\mathbf{k}_{{\rm 1T}})$
for the difference in transverse momentum between the electron and
positron. We use $M_{\mathrm{ee}}$ for the invariant mass of the
electron pair and $\phi$ for the angle between $\mathbf{P}_{{\rm T}}^{\mathrm{ee}}$
and $\mathbf{K}_{{\rm T}}^{\mathrm{ee}}$. We also use $P_{{\rm T}}^{\mathrm{ee}}=|\mathbf{P}_{{\rm T}}^{\mathrm{ee}}|$
and $K_{{\rm T}}^{\mathrm{ee}}=|\mathbf{K}_{{\rm T}}^{\mathrm{ee}}|$
for the lengths of two vectors. 


\section{Theoretical method and setup \label{sec:Theoretical}}

We will use in our calculation the method developed by some of us
for lepton pair photoproduction in the classical field approximation
with the wave packet description of nuclei \citep{Wang:2021kxm,Wang:2022gkd}.
Suppose two identical nuclei $A_{1}$ and $A_{2}$ move in $\pm z$
direction with the velocity $u_{1,2}^{\mu}=\gamma\left(1,0,0,\pm v\right)$
{[}$\gamma=1/\sqrt{1-v^{2}}$ is the Lorentz factor{]} respectively.
Two photons from colliding nuclei produce a lepton pair as $\gamma(p_{1})+\gamma(p_{2})\rightarrow l(k_{1})+l(k_{2})$,
where $p_{1}^{\mu}$ and $p_{2}^{\mu}$ are four-momenta of photons
(photons are not exactly on-shell), and $k_{1}^{\mu}=(E_{k1},\mathbf{k}_{1})$
and $k_{2}^{\mu}=(E_{k2},\mathbf{k}_{2})$ are on-shell four-momenta
leptons. The Born-level total cross section can be written into a
compact form,
\begin{eqnarray}
\sigma & = & \frac{Z^{4}e^{4}}{2\gamma^{4}v^{3}}\int d^{2}\mathbf{b}_{T}d^{2}\mathbf{b}_{1T}d^{2}\mathbf{b}_{2T}\int\frac{d\omega_{1}d^{2}\mathbf{p}_{1T}}{(2\pi)^{3}}\frac{d\omega_{2}d^{2}\mathbf{p}_{2T}}{(2\pi)^{3}}\nonumber \\
 &  & \times\int\frac{d^{2}\mathbf{p}_{1T}^{\prime}}{(2\pi)^{2}}e^{-i\mathbf{b}_{1T}\cdot(\mathbf{p}_{1T}^{\prime}-\mathbf{p}_{1T})}\frac{F^{*}(-\overline{p}_{1}^{\prime2})}{-\overline{p}_{1}^{\prime2}}\frac{F(-\overline{p}_{1}^{2})}{-\overline{p}_{1}^{2}}\nonumber \\
 &  & \times\int\frac{d^{2}\mathbf{p}_{2T}^{\prime}}{(2\pi)^{2}}e^{-i\mathbf{b}_{2T}\cdot(\mathbf{p}_{2T}^{\prime}-\mathbf{p}_{2T})}\frac{F^{*}(-\overline{p}_{2}^{\prime2})}{-\overline{p}_{2}^{\prime2}}\frac{F(-\overline{p}_{2}^{2})}{-\overline{p}_{2}^{2}}\nonumber \\
 &  & \times\int\frac{d^{3}k_{1}}{(2\pi)^{3}2E_{k1}}\frac{d^{3}k_{2}}{(2\pi)^{3}2E_{k2}}(2\pi)^{4}\delta^{(4)}\left(\overline{p}_{1}+\overline{p}_{2}-k_{1}-k_{2}\right)\delta^{(2)}\left(\mathbf{b}_{T}-\mathbf{b}_{1T}+\mathbf{b}_{2T}\right)\nonumber \\
 &  & \times\sum_{\textrm{spin of }l,\overline{l}}\left[u_{1\mu}u_{2\nu}L^{\mu\nu}(\overline{p}_{1},\overline{p}_{2};k_{1},k_{2})\right]\left[u_{1\sigma}u_{2\rho}L^{\sigma\rho*}(\overline{p}_{1}^{\prime},\overline{p}_{2}^{\prime};k_{1},k_{2})\right],\label{eq:cross section}
\end{eqnarray}
where $Z$ is the proton number of the nuclei, $\mathbf{b}_{iT}$
is the transverse position of the photon emission in the nucleus $A_{i}$,
$\mathbf{b}_{T}$ is the impact parameter of colliding nuclei, $\overline{p}_{i}$
and $\overline{p}_{i}^{\prime}$ are photon momenta in the classical
field approximation and defined as ($i=1,2$)
\begin{eqnarray}
\overline{p}_{i}^{\mu} & = & \left(\omega_{i},\boldsymbol{p}_{iT},(-1)^{i+1}\frac{\omega_{i}}{v}\right),\quad\overline{p}_{i}^{\prime\mu}=\left(\omega_{i},\boldsymbol{p}_{iT}^{\prime},(-1)^{i+1}\frac{\omega_{i}}{v}\right),
\end{eqnarray}
satisfying $\overline{p}_{i}\cdot u_{i}=\overline{p}_{i}^{\prime}\cdot u_{i}=0$,
the lepton tensor $L^{\mu\nu}$ is given by 
\begin{eqnarray}
L^{\mu\nu}(p_{1},p_{2};k_{1},k_{2}) & = & -ie^{2}\overline{u}(k_{1})\left[\gamma^{\mu}\frac{\gamma\cdot(k_{1}-p_{1})+m}{(k_{1}-p_{1})^{2}-m^{2}+i\varepsilon}\gamma^{\nu}\right.\nonumber \\
 &  & \left.+\gamma^{\nu}\frac{\gamma\cdot(p_{1}-k_{2})+m}{(p_{1}-k_{2})^{2}-m^{2}+i\varepsilon}\gamma^{\mu}\right]v(k_{2}),\label{eq: Lepton parts}
\end{eqnarray}
and $F(-p^{2})$ is the nuclear charge form factor, Fourier transform
of the nuclear charge density distribution.


\begin{table}
\caption{Parameters for the charge and mass form factors of Ru and Zr and centralities
with corresponding impact parameters. (a) The charge and mass density
distributions from DFT calculation. The centralities and impact parameters
are defined by the mass density distribution; (b) The centralities
and impact parameters are defined by the charge density distribution
determined from DFT calculation. \label{tab:Parameters}}

\centering%
\begin{tabular}{c|c|c|c|ccc|c|c|c|c}
\cline{1-5} \cline{2-5} \cline{3-5} \cline{4-5} \cline{5-5} \cline{7-11} \cline{8-11} \cline{9-11} \cline{10-11} \cline{11-11} 
(a) & $R_{c}$ & $d_{c}$ & $R_{n}$ & $d_{n}$ & $\;$ & {\footnotesize{}Centrality} & 40\% & 60\% & 70\% & 80\%\tabularnewline
\cline{1-5} \cline{2-5} \cline{3-5} \cline{4-5} \cline{5-5} \cline{7-11} \cline{8-11} \cline{9-11} \cline{10-11} \cline{11-11} 
Ru & {\footnotesize{}5.083 fm} & {\footnotesize{}0.477 fm} & {\footnotesize{}5.093 fm} & {\footnotesize{}0.488 fm} &  & {\footnotesize{}Impact parameter} & {\footnotesize{}7.464 fm} & {\footnotesize{}9.143 fm} & {\footnotesize{}9.874 fm} & {\footnotesize{}10.563 fm}\tabularnewline
\cline{1-5} \cline{2-5} \cline{3-5} \cline{4-5} \cline{5-5} \cline{7-11} \cline{8-11} \cline{9-11} \cline{10-11} \cline{11-11} 
Zr & {\footnotesize{}4.977 fm} & {\footnotesize{}0.492 fm} & {\footnotesize{}5.022 fm} & {\footnotesize{}0.538 fm} &  & {\footnotesize{}Impact parameter} & {\footnotesize{}7.615 fm} & {\footnotesize{}9.326 fm} & {\footnotesize{}10.073fm} & {\footnotesize{}10.780 fm}\tabularnewline
\cline{1-5} \cline{2-5} \cline{3-5} \cline{4-5} \cline{5-5} \cline{7-11} \cline{8-11} \cline{9-11} \cline{10-11} \cline{11-11} 
\multicolumn{1}{c}{} & \multicolumn{1}{c}{} & \multicolumn{1}{c}{} & \multicolumn{1}{c}{} &  &  & \multicolumn{1}{c}{} & \multicolumn{1}{c}{} & \multicolumn{1}{c}{} & \multicolumn{1}{c}{} & \tabularnewline
\cline{1-5} \cline{2-5} \cline{3-5} \cline{4-5} \cline{5-5} \cline{7-11} \cline{8-11} \cline{9-11} \cline{10-11} \cline{11-11} 
(b) & $R_{c}$ & $d_{c}$ & $R_{n}$ & $d_{n}$ &  & {\footnotesize{}Centrality} & 40\% & 60\% & 70\% & 80\%\tabularnewline
\cline{1-5} \cline{2-5} \cline{3-5} \cline{4-5} \cline{5-5} \cline{7-11} \cline{8-11} \cline{9-11} \cline{10-11} \cline{11-11} 
Ru & {\footnotesize{}5.083 fm} & {\footnotesize{}0.477 fm} & $R_{c}^{\mathrm{Ru}}$ & $d_{c}^{\mathrm{Ru}}$ &  & {\footnotesize{}Impact parameter} & {\footnotesize{}7.406 fm} & {\footnotesize{}9.070 fm} & {\footnotesize{}9.797 fm} & {\footnotesize{}10.479 fm}\tabularnewline
\cline{1-5} \cline{2-5} \cline{3-5} \cline{4-5} \cline{5-5} \cline{7-11} \cline{8-11} \cline{9-11} \cline{10-11} \cline{11-11} 
Zr & {\footnotesize{}4.977 fm} & {\footnotesize{}0.492 fm} & $R_{c}^{\mathrm{Zr}}$ & $d_{c}^{\mathrm{Zr}}$ &  & {\footnotesize{}Impact parameter} & {\footnotesize{}7.373 fm} & {\footnotesize{}9.030 fm} & {\footnotesize{}9.754 fm} & {\footnotesize{}10.434 fm}\tabularnewline
\cline{1-5} \cline{2-5} \cline{3-5} \cline{4-5} \cline{5-5} \cline{7-11} \cline{8-11} \cline{9-11} \cline{10-11} \cline{11-11} 
\end{tabular}
\end{table}

The density distributions for the nuclear charge (proton) and mass
(nucleon) are obtained from the state-of-art DFT calculation. Same
as Ref.~\citep{Xu:2021vpn,Xu:2021qjw}, we parameterize the charge
or mass density distributions with the Woods-Saxon (WS) distribution,
\begin{equation}
\rho_{i}(\mathbf{r})\equiv\frac{C_{i}}{1+\exp[(|\mathbf{r}|-R_{i})/d_{i}]},\label{eq:WS}
\end{equation}
by matching $\left\langle r\right\rangle $ and $\left\langle r^{2}\right\rangle $.
Here $i=c,n$ denotes the charge and mass density distribution respectively,
and $R_{i}$, $d_{i}$, $C_{i}$ are the corresponding radius, skin
depth and normalization factor, respectively. In the calculation,
we use the charge density distribution for the cross section in Eq.~(\ref{eq:cross section}),
so $F(-p^{2})$ is the nuclear charge form factor defined as $F_{c}(\mathbf{k}^{2})=\int d^{3}re^{i\mathbf{k}\cdot\mathbf{r}}\rho_{c}(\mathbf{r})$.
We use the mass density distribution $\rho_{n}(\mathbf{r})$ to determine
the impact parameter $\mathbf{b}_{T}$ through the centrality of collisions.


The charge and mass density distributions give sizable difference
in the multiplicity and thus centrality in isobar collisions~\citep{Deng:2016knn,Xu:2017zcn,Li:2018oec}.
The WS parameters for the charge and mass density distributions and
the centralities with corresponding impact parameters are listed in
Tab.~\ref{tab:Parameters}. In Tab.~\ref{tab:Parameters}(a) the
centralities with corresponding impact parameters are defined by the
mass density distribution which has different WS parameters from the
charge density distribution. We will label the numerical results in
this case as ``(a) DFT''. For comparison, we also use the charge
density distribution as the mass density distribution to define the
centrality as shown in Tab.~\ref{tab:Parameters}(b). We will label
the numerical results in this case as ``(b) charge$=$mass''. The
impact parameters and centralities are calculated by an optical Glauber
model as in Ref.~\citep{Miller:2007ri} with $\sigma_{{\rm NN}}=$42
mb at $\sqrt{s_{{\rm NN}}}=$200 GeV. We have checked that parameters
used in Ref.~\citep{Deng:2016knn} belong to the case ``(b) charge$=$mass''.
This indicates that the lepton pair photoproduction in isobar collisions
can reflect the information about nuclear structure of colliding nuclei.


In principle, the deformation of isobar nuclei~\citep{STAR:2021mii,zhang:2021kxj}
can also change the profile of classical photon fields and introduce
a higher order correction to the differential cross sections. However,
a systematic study of the deformation effect may require computing
the differential cross section (\ref{eq:cross section}) with the
classical photon fields on the event-by-event basis, which is challenging.
Therefore, as a first attempt, we do not consider the deformation
effect in the current study.


We follow STAR experiments~\citep{STAR:2018ldd,STAR:2019wlg,Zhou:2022gbh}
to set that the transverse momentum of the electron or positron is
greater than 200 MeV. The rapidity range of $e^{+}e^{-}$ pair and
pseudo-rapidity range of a single electron or positron are set to
$[-1,1]$. To deal with the high dimensional integration in Eq.~(\ref{eq:cross section}),
we implement the ZMCintegral package~\citep{Wu:2019tsf,Zhang:2019nhd}
(also see Ref.~\citep{Zhang:2019uor,Zhang:2022lje} for other applications
of the package).


\section{Transverse momentum, invariant mass and azimuthal angle distributions
\label{sec:distriubtion}}

In this section, we present numerical results for transverse momentum,
invariant mass and azimuthal angle distributions for $e^{+}e^{-}$
in Ru+Ru and Zr+Zr collisions at $\sqrt{s_{\mathrm{NN}}}=$ 200 GeV.
Without the difference in the nuclear structure, the ratio of the
differential cross section in Ru+Ru collisions to Zr+Zr collisions
at the same collision energy should only depend on the number of protons,
i.e. $(44/40)^{4}$, as in Eq.~(\ref{eq:cross section}). However,
we will show that the difference in nuclear structure in terms of
proton and nucleon distributions in Ru and Zr will make a difference
in the cross section.


\begin{figure}
\centering\includegraphics[scale=0.45]{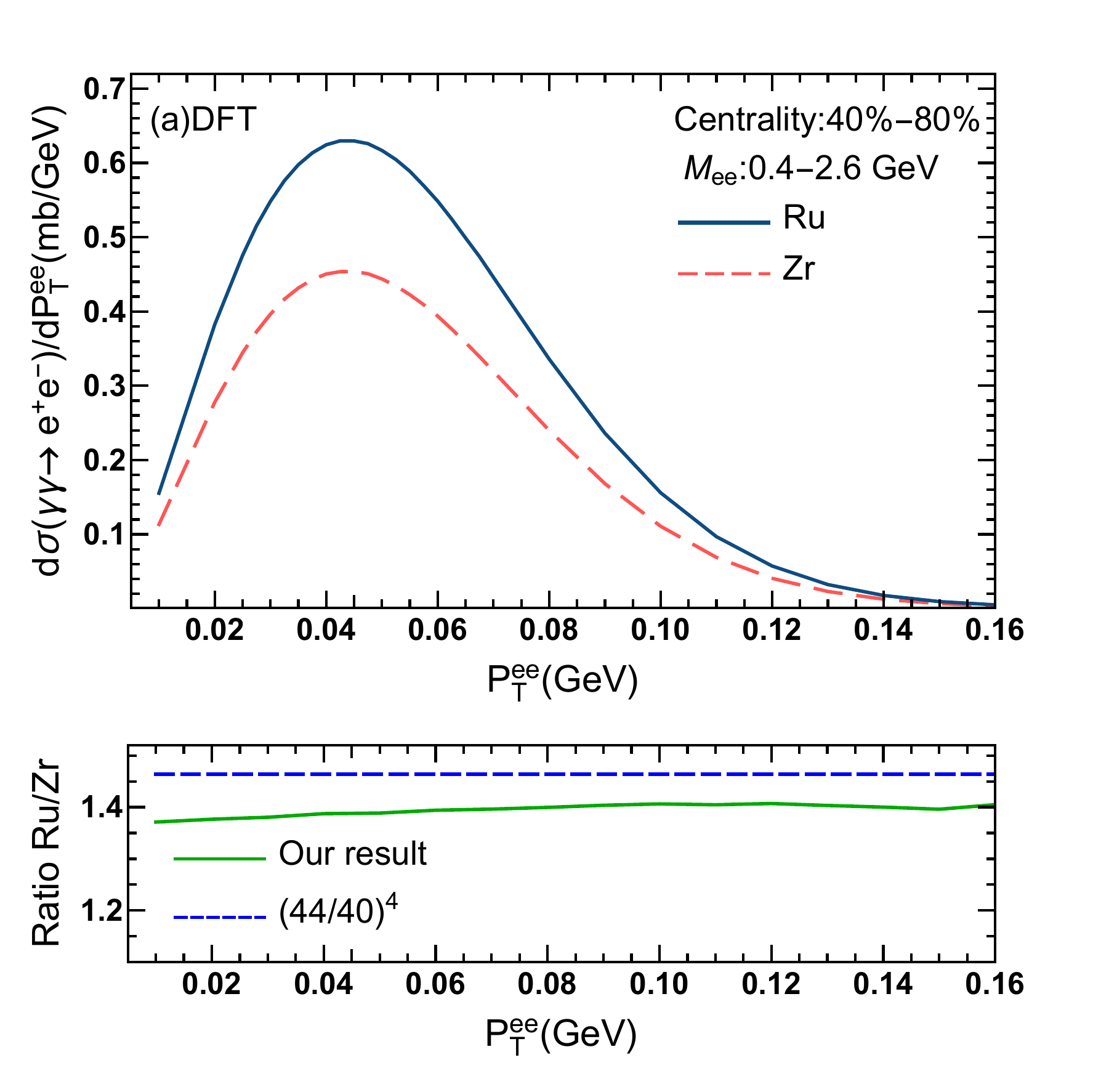}\includegraphics[scale=0.45]{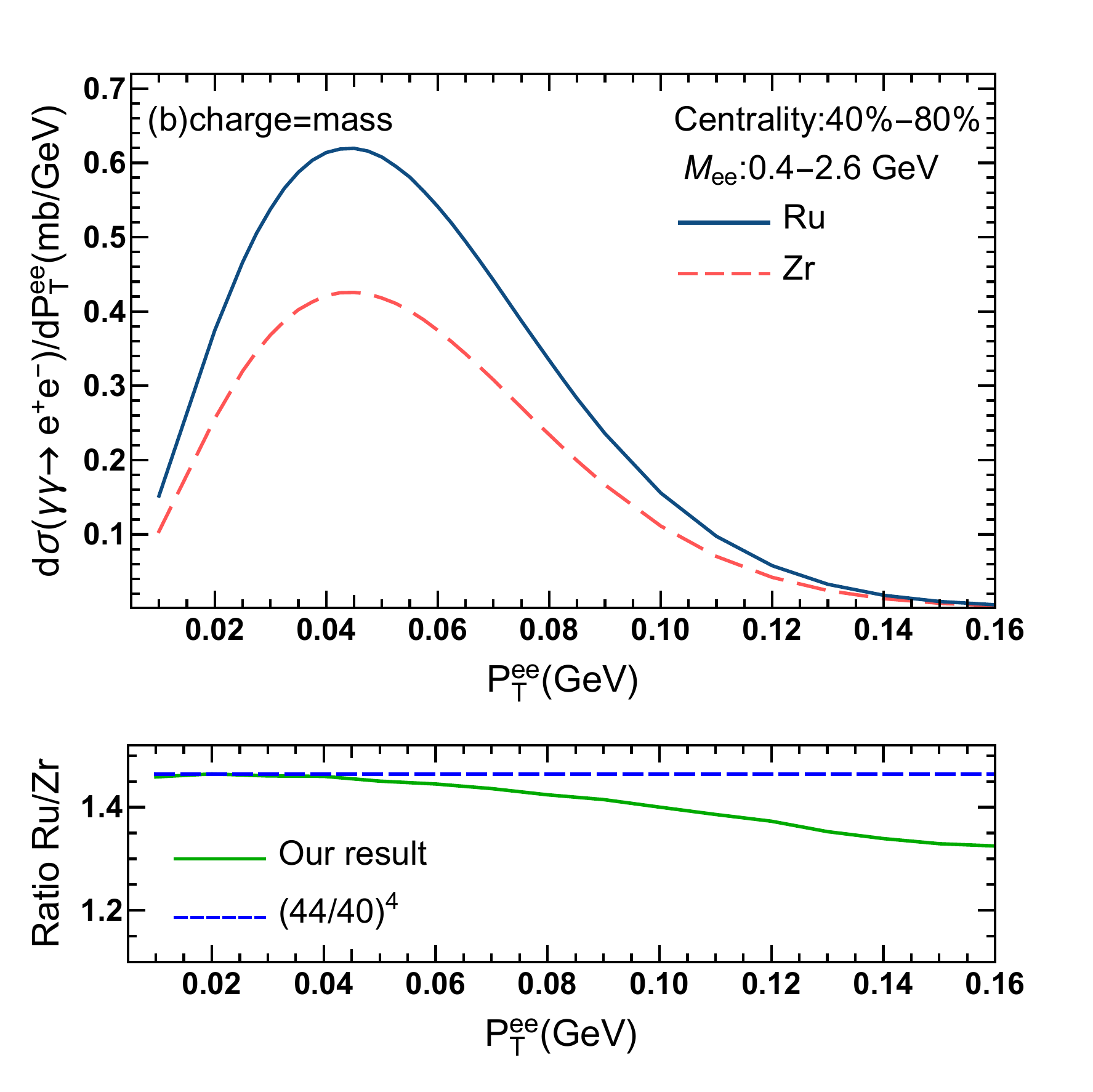}\caption{Differential cross sections in Ru+Ru and Zr+Zr collisions at 40-80\%
centrality as functions of $P_{{\rm T}}^{{\rm ee}}$. The subfigure
(a) and (b) correspond to the set of parameters in Tab.~\ref{tab:Parameters}(a)
and Tab.~\ref{tab:Parameters}(b) from the DFT calculation, respectively.
In (b), we use the short-hand notation \textquotedblleft charge=mass\textquotedblright{}
for the case in Tab.~\ref{tab:Parameters}(b) in which the nuclear
mass density distribution is chosen to be the same as the charge density
distribution. In (a), the blue-solid and red-dashed lines are the
results for Ru+Ru and Zr+Zr collisions, respectively. In (b), the
green-solid line is the ratio of the differential cross section in
Ru+Ru collisions to Zr+Zr collisions. The range of $M_{{\rm ee}}$
is set to {[}0.4,2.6{]} GeV. \label{fig:Pt}}
\end{figure}

In Fig.~\ref{fig:Pt}, we present the $P_{{\rm T}}^{{\rm ee}}$ distribution
at 200 GeV in the Ru+Ru and Zr+Zr collisions. We observe that $d\sigma_{{\rm Ru+Ru}}/dP_{{\rm T}}^{{\rm ee}}$
is always larger than $d\sigma_{{\rm Zr+Zr}}/dP_{{\rm T}}^{{\rm ee}}$
due to different charges in Ru and Zr. The shapes of $d\sigma_{{\rm Ru+Ru}}/dP_{{\rm T}}^{{\rm ee}}$
and $d\sigma_{{\rm Zr+Zr}}/dP_{{\rm T}}^{{\rm ee}}$ are similar:
the spectra have peaks at $P_{{\rm T}}^{{\rm ee}}\sim$45 MeV and
drop rapidity when $P_{{\rm T}}^{{\rm ee}}>$45 MeV. We also show
the ratio of $d\sigma_{{\rm Ru+Ru}}/dP_{{\rm T}}^{{\rm ee}}$ to $d\sigma_{{\rm Zr+Zr}}/dP_{{\rm T}}^{{\rm ee}}$.
The ratio is expected to be $(Z_{\mathrm{Ru}}/Z_{\mathrm{Zr}})^{4}=(44/40)^{4}$
but there is a little difference. As shown in Tab.~\ref{tab:Parameters},
such a difference is mainly due to the difference in impact parameters
at given centralities and the nuclear charge form factors in Ru+Ru
and Zr+Zr collisions. We see that the charge and mass distributions
can lead to a difference between (a) and (b) in the range $P_{{\rm T}}^{{\rm ee}}\leq$40
MeV. For case (a) {[}Fig.~\ref{fig:Pt}(a){]} in which one distinguishes
nuclear charge and mass density distributions, the ratio is always
smaller than $(44/40)^{4}$, while for case (b) in which the nuclear
mass density distribution is set to be equal to the charge density
distribution, it is very close to $(44/40)^{4}$ at $P_{{\rm T}}^{{\rm ee}}\leq$40
MeV. So the fine nuclear structure can affect the dilepton spectra
and therefore can be directly measured in experiments. The ratio shown
in Fig.~\ref{fig:Pt}(a) is one of our main prediction in this study.


\begin{figure}
\centering\includegraphics[scale=0.6]{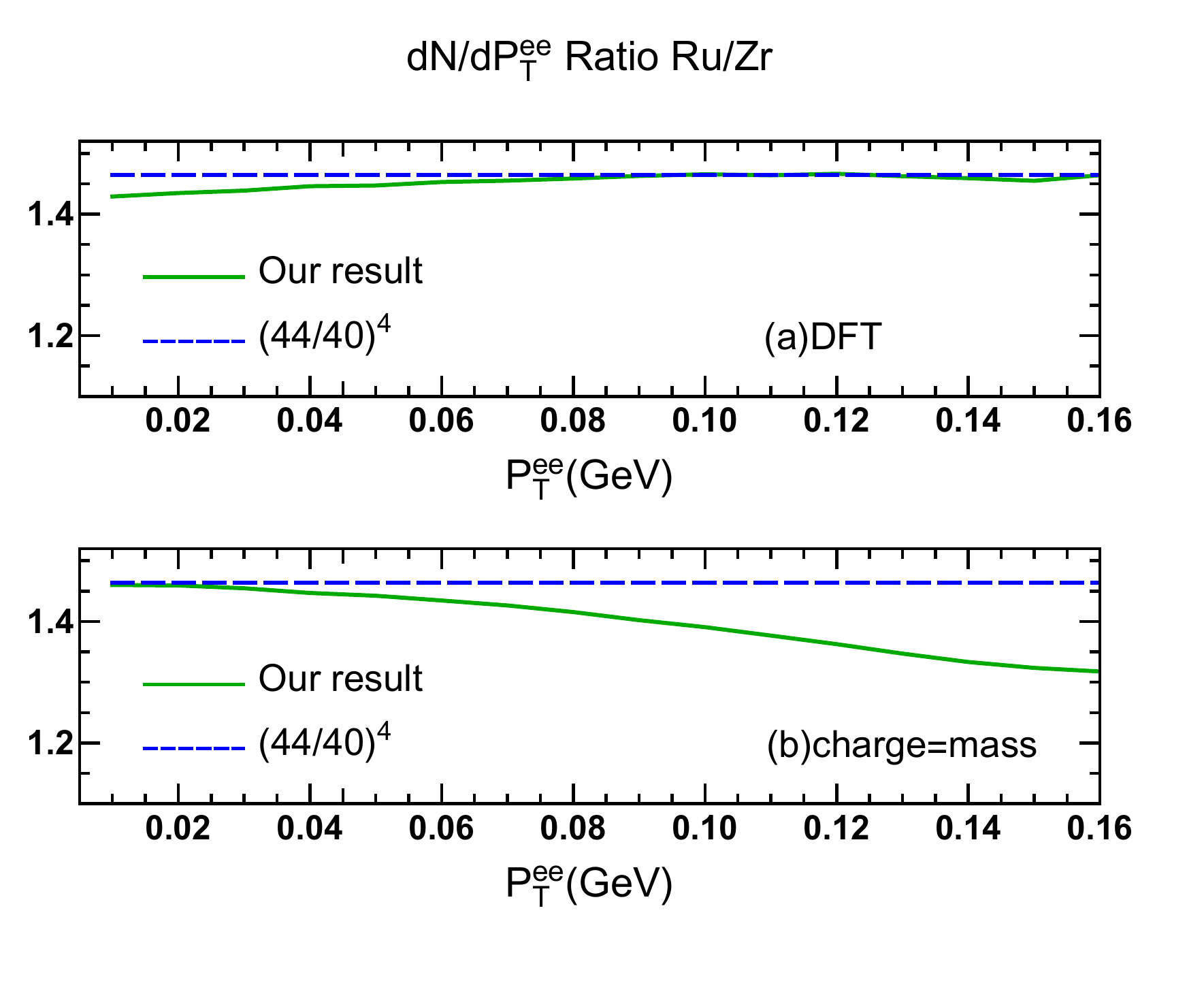}\caption{The ratio of $dN/dP_{{\rm T}}^{{\rm ee}}$ in Ru+Ru to Zr+Zr collisions
as functions of $P_{{\rm T}}^{{\rm ee}}$ at 40-80\% centrality. The
definition of green-solid and blue-solid lines is the same as in Fig.
\ref{fig:Pt}. \label{fig:Pt-2}}
\end{figure}

We also calculate the ratio of $dN_{{\rm Ru+Ru}}/dP_{{\rm T}}^{{\rm ee}}$
to $dN_{{\rm Zr+Zr}}/dP_{{\rm T}}^{{\rm ee}}$ in Fig.~\ref{fig:Pt-2},
where the differential yield $dN$ is proportional to $d\sigma$ as
\begin{eqnarray}
dN & = & \frac{d\sigma}{\pi(b_{{\rm T;max}}^{2}-b_{{\rm T;min}}^{2})}.
\end{eqnarray}
We observe that in case (a) the ratio of $dN/dP_{{\rm T}}^{{\rm ee}}$
for Ru+Ru and Zr+Zr collisions is slightly smaller than $(44/40)^{4}$
at very small $P_{{\rm T}}^{{\rm ee}}$ ($P_{{\rm T}}^{{\rm ee}}\leq$80
MeV) by about 3\%. The ratio in case (b) is similar to Fig.~\ref{fig:Pt}
(b).

\begin{figure}
\centering\includegraphics[scale=0.45]{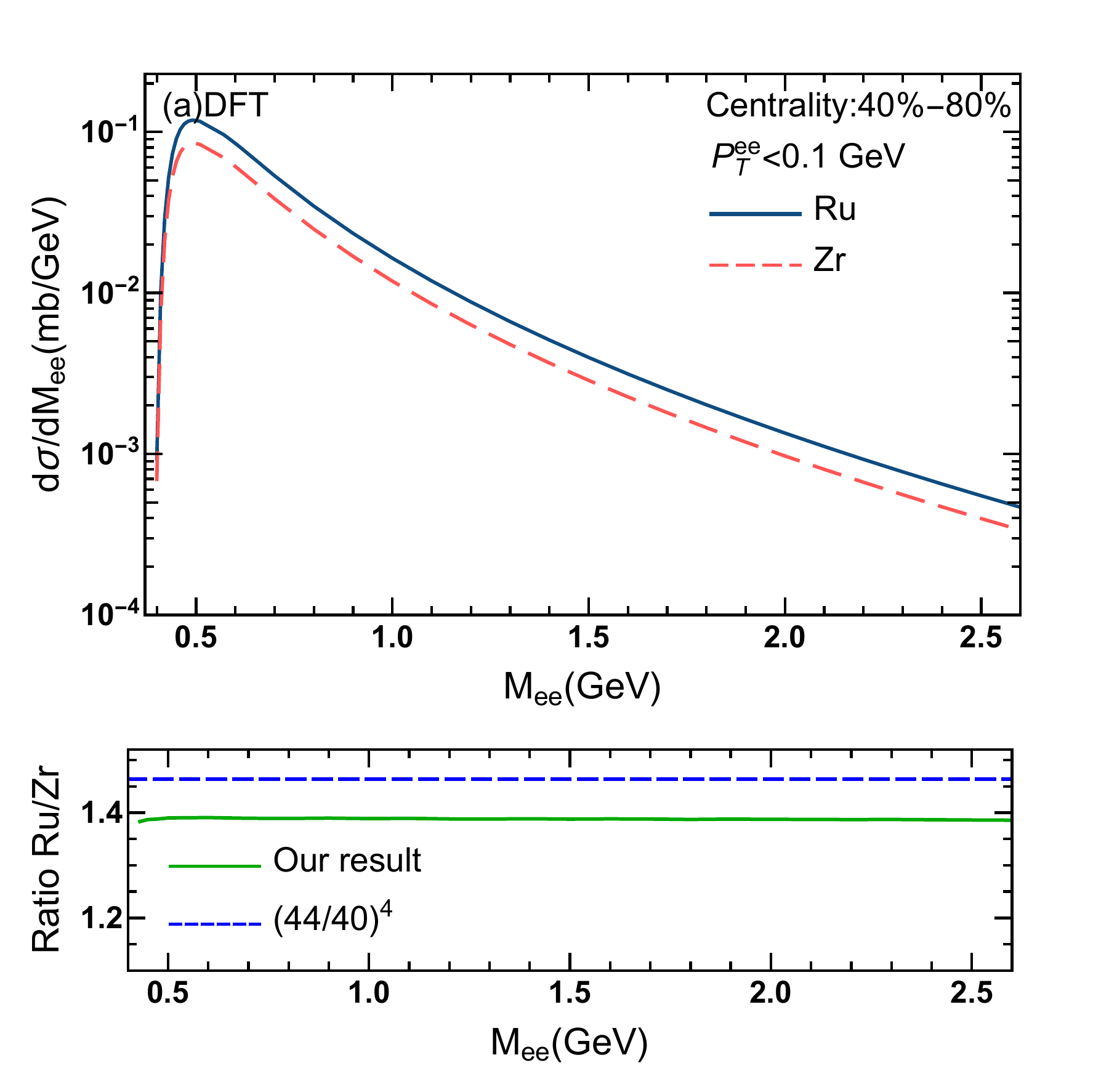}\includegraphics[scale=0.45]{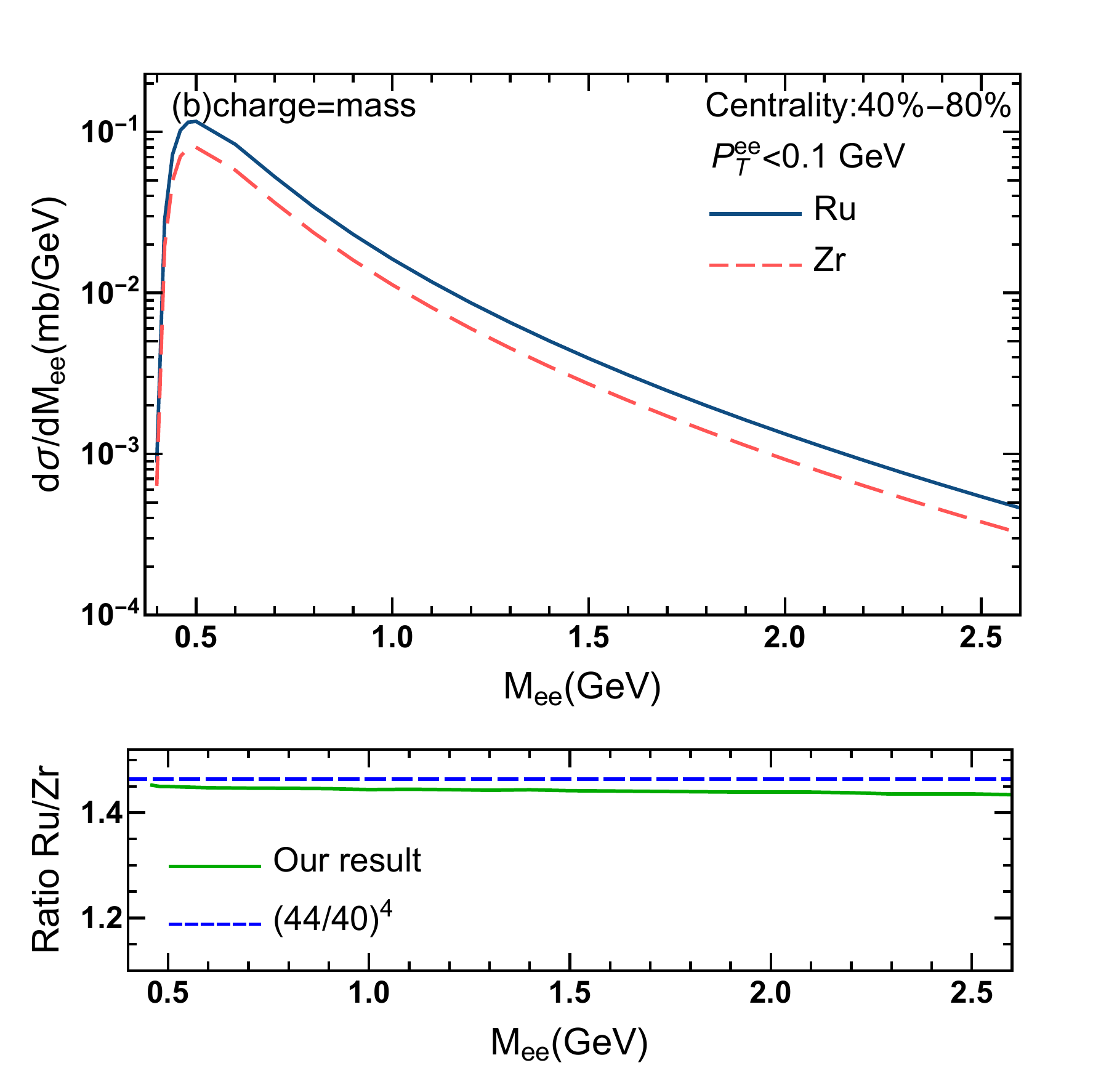}\caption{Differential cross sections of $e^{+}e^{-}$ in 40-80\% centrality
as functions of $M_{{\rm ee}}$. The parameters used in (a) and (b)
are the same as in Fig. (\ref{fig:Pt}). The definition of color-dashed
or color-solid lines are the same as Fig. (\ref{fig:Pt}). The range
of $P_{{\rm T}}^{{\rm ee}}$ is chosen as $P_{{\rm T}}^{{\rm ee}}<$0.1
GeV. \label{fig:mee}}
\end{figure}

The results for the differential cross sections as functions of the
invariant mass of $e^{+}e^{-}$ are shown in Fig.~\ref{fig:mee}.
The behavior of invariant mass spectra in Ru+Ru and Zr+Zr collisions
is similar to Au+Au collisions~\citep{Wang:2021kxm,Wang:2022gkd,STAR:2019wlg,Zha:2018ywo,Zha:2018tlq}.
The spectra have peaks at $M_{{\rm ee}}\sim$0.5 GeV and then decreases
exponentially with increasing $M_{{\rm ee}}$. In case (a) with the
parameters in Tab.~\ref{tab:Parameters}(a), the ratio of $d\sigma_{{\rm Ru+Ru}}/dM_{{\rm ee}}$
to $d\sigma_{{\rm Zr+Zr}}/dM_{{\rm ee}}$ is always smaller than $(44/40)^{4}$,
while the ratio in case (b) with the parameters in Tab.~\ref{tab:Parameters}(b)
is almost equal to $(44/40)^{4}$.

\begin{figure}
\centering\includegraphics[scale=0.45]{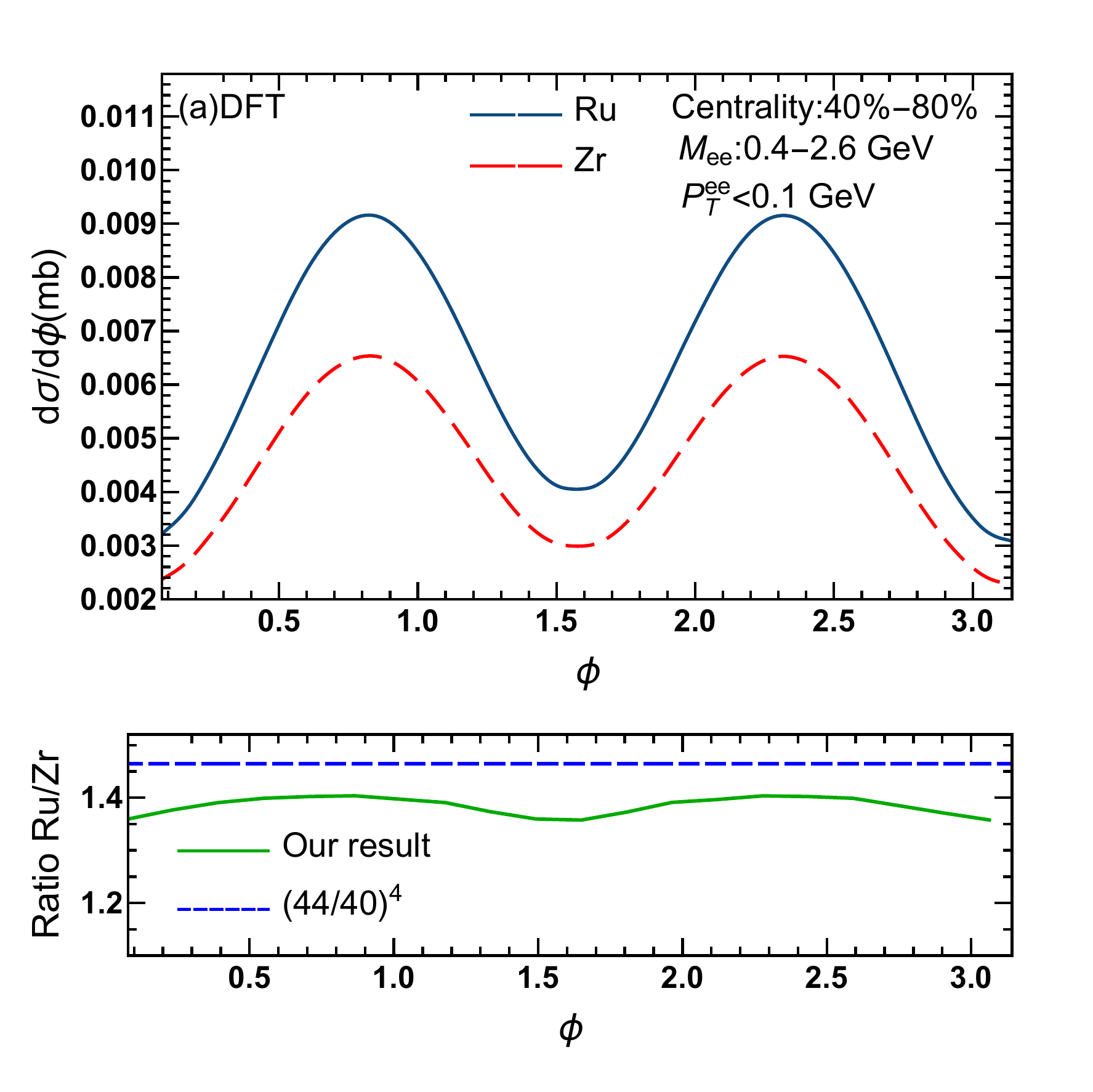} \includegraphics[scale=0.45]{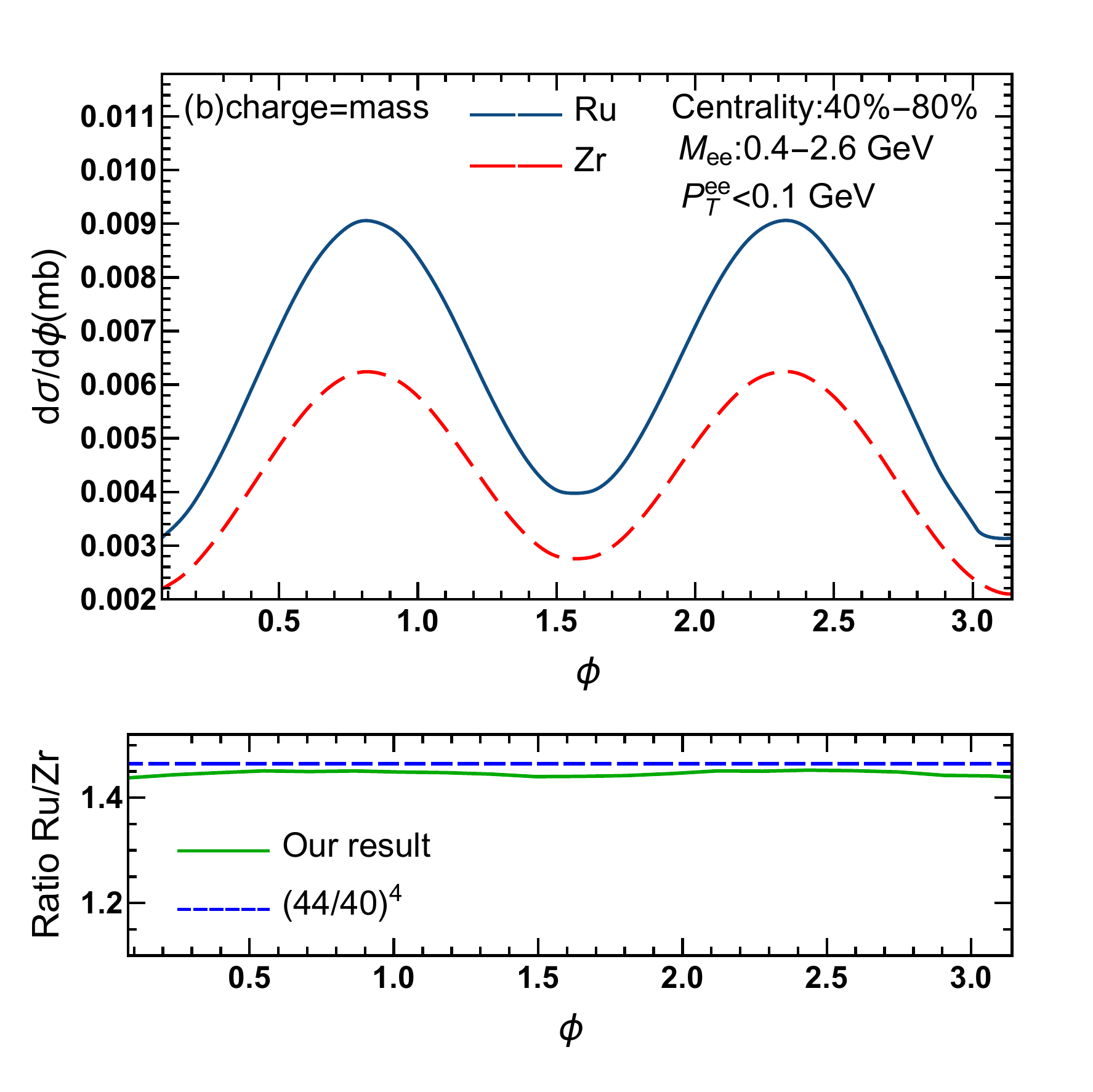}\caption{Differential cross sections in Ru+Ru and Zr+Zr collisions as functions
of $\phi$ at 40-80\% centrality. The definition of all color-solid
and -dashed lines is the same as in Fig. \ref{fig:Pt}. The ranges
$M_{{\rm ee}}$ and $P_{{\rm T}}^{{\rm ee}}$ are chosen as $M_{{\rm ee}}\in${[}0.4,2.6{]}
GeV and $P_{{\rm T}}^{{\rm ee}}<$0.1 GeV. \label{fig:cos}}
\end{figure}

The azimuthal angle distributions at 40-80\% centrality in isobar
collisions at 200 GeV are shown in Fig.~\ref{fig:cos}. Our results
show a modulation of $-\cos(4\phi)$ in Ru+Ru and Zr+Zr collisions
with two sets of parameters in Tab.~\ref{tab:Parameters}. Such a
modulation is related to the linear polarization of incident photons~\citep{Li:2019sin,Li:2019yzy,Wang:2022gkd},
see Refs.~\citep{Akcakaya:2012si,Schafer:2012yx,Boer:2017xpy,Dominguez:2011br,Metz:2011wb,Hagiwara:2021qev}
for similar effects in QCD. The ratio of $d\sigma_{{\rm Ru+Ru}}/d\phi$
to $d\sigma_{{\rm Zr+Zr}}/d\phi$ is always smaller than $(44/40)^{4}$
in case (a), while it agrees with $(44/40)^{4}$ in case (b).

We see that the ratios of differential cross sections in Ru+Ru collisions
to Zr+Zr collisions as functions of $P_{{\rm T}}^{{\rm ee}}$, $M_{{\rm ee}}$
and $\phi$ in case (a) are always smaller than $(44/40)^{4}$. So
it provides a way of measuring the difference between the charge and
mass distributions in isobar nuclei through $P_{{\rm T}}^{{\rm ee}},$
$M_{{\rm ee}}$ and $\phi$ spectra of lepton pairs in isobar collisions.


\section{Charge and centrality dependence \label{sec:Charge-and-centrality}}

In this section, we study the charge and centrality dependence of
cross sections in isobar collisions. We have presented the spectra
of $P_{{\rm T}}^{{\rm ee}}$, $M_{{\rm ee}}$ and $\phi$ for di-electrons
by using the set of parameters in case (a) and (b) in the previous
section. In this section we will focus on case (a) in which the nuclear
charge and mass density distributions are distinguished. In general,
the deformation of colliding nuclei can have an effect on the spectra,
but we do not consider it here and leave it for a future study.

\begin{figure}
\centering\includegraphics[scale=0.35]{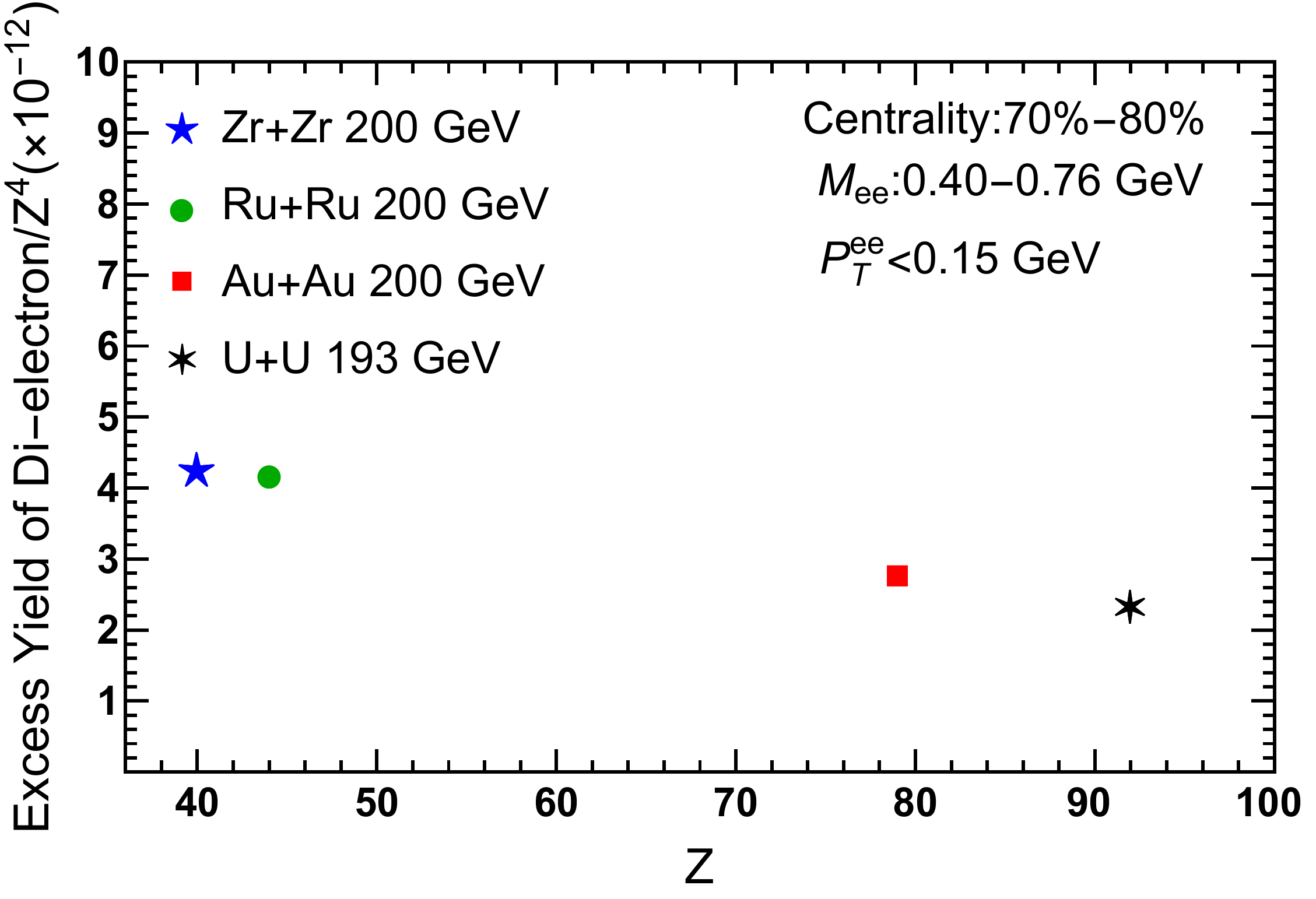}\caption{The charge dependence of the integrated excess yields scaled by $Z^{4}$
in 70-80\% centrality in Ru+Ru, Zr+Zr, Au+Au and U+U collisions. The
ranges of $M_{{\rm ee}}$ and $P_{{\rm T}}^{{\rm ee}}$ are $M_{{\rm ee}}\in${[}0.4,0.76{]}
GeV and $P_{{\rm T}}^{{\rm ee}}<$0.15 GeV. \label{fig:Charge dependence}}
\end{figure}

In Fig.~\ref{fig:Charge dependence}, we present the charge dependence
of the integrated excess yield $N$ scaled by $Z^{4}$, which is defined
as the cross section in a specific range of impact parameters, 
\begin{equation}
N=\frac{\int_{b_{{\rm T;min}}}^{b_{{\rm T;max}}}d\sigma}{\pi(b_{{\rm T;max}}^{2}-b_{{\rm T;min}}^{2})}.
\end{equation}
The result shows that the integrated excess yield scaled by $Z^{4}$
decreases with $Z$, which is consistent with Ref.\citep{STAR:2018ldd,Zha:2018tlq}.
Such a behavior is caused by different charge and mass distributions
in different nuclei. For a given centrality, the ranges of impact
parameters of different nuclei are different due to their mass distributions.

The centrality dependence of the integrated excess yield is summarized
in Tab.~\ref{tab:centrality}. We see that the ratio of the integrated
excess yield in Ru+Ru collisions to Zr+Zr collisions is smaller than
$(44/40)^{4}\approx1.4641$, which is consistent with the results
in the previous section. The ratio increases slightly with the impact
parameter.

\begin{table}
\caption{The centrality dependence of the integrated excess yield at 200 GeV
in Ru+Ru and Zr+Zr collisions. The range of $M_{{\rm ee}}$ and $P_{{\rm T}}^{{\rm ee}}$
are set to $M_{{\rm ee}}\in${[}0.4,2.6{]} GeV and $P_{{\rm T}}^{{\rm ee}}<$0.1
GeV. \label{tab:centrality}}

\centering%
\begin{tabular}{c|c|c|c}
\hline 
 & Ru & Zr & ratio Ru/Zr\tabularnewline
\hline 
\hline 
40-60\% & $2.328\times10^{-5}$ & $1.615\times10^{-5}$ & $1.441$\tabularnewline
\hline 
60-70\% & $2.245\times10^{-5}$ & $1.549\times10^{-5}$ & $1.449$\tabularnewline
\hline 
70-80\% & $2.178\times10^{-5}$ & $1.495\times10^{-5}$ & $1.457$\tabularnewline
\hline 
\end{tabular}
\end{table}

\begin{figure}
\centering\includegraphics[scale=0.35]{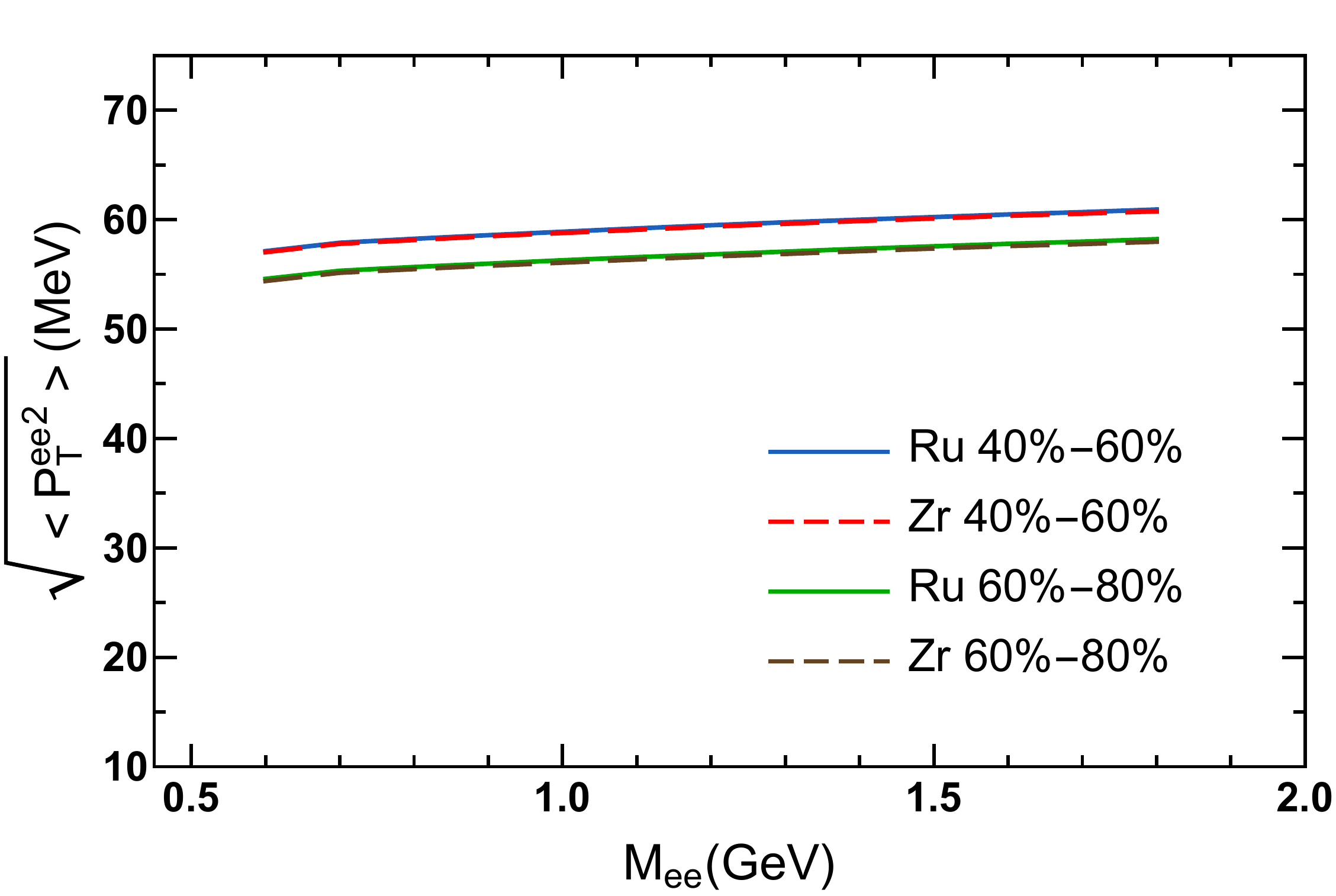}\caption{The average transverse momenta of di-electrons as functions of $M_{{\rm ee}}$
in 40-60\% and 60-80\% centrality ranges. The blue and green solid
lines stand for 40-60\% and 60-80\% centralities in Ru+Ru collisions
respectively. The red and brown dashed lines stand for the same centralities
in Zr+Zr collisions respectively. We choose $P_{{\rm T}}^{{\rm ee}}<$0.1
GeV. \label{fig:Pt broadening}}
\end{figure}

In Fig.~\ref{fig:Pt broadening}, we show $\sqrt{\left\langle (P_{{\rm T}}^{{\rm ee}})^{2}\right\rangle }$
as functions of $M_{{\rm ee}}$ in 40-60\% and 60-80\% centrality
ranges in Ru+Ru and Zr+Zr collisions. We see that $\sqrt{\left\langle (P_{{\rm T}}^{{\rm ee}})^{2}\right\rangle }$
increases with increasing $M_{{\rm ee}}$ or decreasing impact parameters.
The trend agrees with the previous work in Ref.~\citep{Wang:2022gkd}
by some of us in Au+Au collisions. In contrast to $d\sigma/dP_{{\rm T}}^{{\rm ee}}$
in Fig.~\ref{fig:Pt}, $\sqrt{\left\langle (P_{{\rm T}}^{{\rm ee}})^{2}\right\rangle }$
as functions of $M_{{\rm ee}}$ in Ru+Ru or Zr+Zr collisions are almost
the same.


\section{Conclusion \label{sec:Conclusion}}

We have investigated the photoproduction of di-electrons in peripheral
isobar collisions based on the method developed in previous works
by some of us~\citep{Wang:2021kxm,Wang:2022gkd}. The nuclear mass
and charge density distributions are described by Woods-Saxon distributions
with the parameters determined from the DFT calculation. The charge
density distribution gives the nuclear charge form factor in the photoproduction
cross section for di-electrons. The centrality corresponds to the
impact parameter in a Glauber model and is determined through the
nuclear mass density distribution. According to the DFT, the nuclear
mass density distribution is different from the nuclear charge density
distribution in general. For comparison, we also consider the approximation
that the mass density distribution is taken as equal to the charge
density distribution.


We calculated the spectra of $P_{{\rm T}}^{{\rm ee}}$, $M_{{\rm ee}}$
and $\phi$ at 40-80\% centrality in Ru+Ru and Zr+Zr collisions at
200 GeV. We take the ratio of these spectra in Ru+Ru collisions to
Zr+Zr collisions. The results are presented in Fig.~\ref{fig:Pt},
Fig.~\ref{fig:Pt-2}, Fig.~\ref{fig:mee} and Fig.~\ref{fig:cos}
which show the effect arising from the difference between the mass
and charge density distributions. If one does not distinguish the
charge and mass density distributions, these ratios are close to $(44/40)^{4}$,
the ratio of the fourth power of the proton number. If one does distinguish
them, these ratios are generally smaller than $(44/40)^{4}$ by a
few percent. This is the main result of our paper.


We also calculated the charge and centrality dependence of integrated
excess yield in isobar collisions. We find that it decreases with
growing $Z$ as shown in Fig.~\ref{fig:Charge dependence}. Such
a behavior reflects different charge and mass distributions in different
nuclei. We present in Fig.~\ref{fig:Pt broadening} the spectra of
$\sqrt{\left\langle (P_{{\rm T}}^{{\rm ee}})^{2}\right\rangle }$
as functions of $M_{{\rm ee}}$ in 40-60\% and 60-80\% centrality
ranges in Ru+Ru and Zr+Zr collisions. The $M_{{\rm ee}}$ spectra
of $\sqrt{\left\langle (P_{{\rm T}}^{{\rm ee}})^{2}\right\rangle }$
are found to be similar to Au+Au collisions~\citep{Wang:2022gkd}
and is almost the same in Ru+Ru and Zr+Zr collisions.

With above results, we conclude that the photoproduction of lepton
pairs in isobar collisions may provide a new way to probe the nuclear
structure. We note that nuclear deformation effects are neglected
in this paper and may have effects on above observables, which worths
a future study.

\begin{acknowledgments}
We would like to thank Fuqiang Wang, Xiao-Feng Wang, Bo-Wen Xiao and
Jian Zhou for helpful discussion. This work is partly supported by
National Natural Science Foundation of China (NSFC) under Grants No.
11909059, 12035006, 12075235, 12135011 and 12275082. 
\end{acknowledgments}

\bibliographystyle{h-physrev}
\bibliography{UPCisobar}

\end{document}